\documentclass[aps,pre,twocolumn,,groupedaddress]{revtex4}
\usepackage{graphicx,color,subfigure}

\begin{document}
\title{Structure and dynamics of model colloidal clusters with short-range attractions}
\author{Robert S. Hoy}
\email{rshoy@usf.edu}
\affiliation{Department of Physics, University of South Florida, Tampa, FL, 33620}
\date{\today}

\begin{abstract}
We examine the structure and dynamics of small isolated $N$-particle clusters interacting via short-ranged Morse potentials.  
``Ideally preprared ensembles'' obtained via exact enumeration studies of sticky hard sphere packings serve as reference states allowing us to identify key statistical-geometrical properties and to quantitatively characterize how nonequilibrium ensembles prepared by thermal quenches at different rates $\dot{T}$ differ from their equilibrium counterparts. 
Studies of equilibrium dynamics show nontrival temperature dependence: nonexponential relaxation indicates both glassy dynamics and differing stabilities of degenerate clusters with different structures.
Our results should be useful for extending recent experimental studies of small colloidal clusters to examine both equilibrium relaxation dynamics at fixed $T$ and a variety of nonequilibrium phenomena.
\end{abstract}
\maketitle

\section{Introduction}
\label{sec:intro}

Understanding how varying the shape and strength of a pair potential affects the energy landscape and dynamics of systems composed of several particles interacting via that potential lies at the heart of theoretical cluster physics \cite{wales04}.
Variable-shape potentials are of particular utility in understanding common features of apparently disparate systems.
For example, varying the dimensionless range parameter $\alpha D$ of the Morse potential
\begin{equation}
U_{\rm Morse}(\alpha; r) = \epsilon\left[\exp(-2\alpha(r-D)) - 2\exp(-\alpha(r-D))\right],
\label{eq:morsepot}
\end{equation}
yields accurate models for clusters formed by constituents ranging from alkali-metal atoms to buckyballs to micron-sized colloids \cite{calvo12}.
Studies of colloidal clusters are particularly valuable in this context since individual particle positions can be tracked.
Most valuable are ``model'' systems with precisely controllable interparticle interactions and cluster size $N$.
These systems are a veritable playground for studies of few-body statistical mechanics, and can (through the universality evident in cluster physics) provide insights into the behavior of their more microscopic counterparts.

Manoharan and collaborators have recently attracted great interest by characterizing the structure and dynamics \cite{meng10,perry12} of colloids interacting via hard-core-like repulsive and (variably) short-ranged attractive interactions.
While published experimental studies and related theoretical modeling  \cite{meng10,arkus09,arkus11,hoy10,hoy12,wales10,calvo12,khan12,klix13,morgan14,holmes14}  of these systems have focused on equilibrium phenomena, rapid advances in experimental particle-tracking techniques \cite{perry12,fung12,fung13} suggest that much of their \textit{nonequilibrium} physics may soon become experimentally observable.
For example, the room-temperature transition rate between the two degenerate ground-state clusters (GSC) of $N=6$ particles is of order $10^{-3}-10^{0} s^{-1}$.
Since their longest relaxation times should increase dramatically with increasing $N$ and decreasing temperature $T$, it seems plausible that these model colloidal systems could soon be utilized for fundamental studies of \textit{nonequilibrium} few-body statistical mechanics.

In this paper, we provide theoretical guidance for such studies by elucidating the statistical-geometrical properties and several key equilibrium and nonequilibrium phenomena in small $(N \leq 13)$ clusters that mimic the systems studied in experiments  \cite{malins09,meng10,perry12}.
First we perform exact-enumeration studies that extend the work of Refs.\ \cite{arkus09,hoy10,arkus11,hoy12} by obtaining all \textit{minimally} \cite{jacobs95} mechanically stable packings of $N \leq 13$ sticky hard spheres.  
Then we use the ``ideally prepared ensembles'' of ground-state clusters generated by these studies as initial conditions for molecular dynamics simulations of $N$-particle model colloidal clusters.
These simulations focus on identifying noteworthy features in their equilibrium relaxation dynamics and their preparation-protocol-dependent, nonequilibrium structure that should be observable in particle-tracking experiments.

Our key results are that:\ \textbf{(i)} the fraction of ``off-pathway nuclei'' that are mechanically stable yet incompatible with close-packed crystallization grows rapidly with increasing $N$; \textbf{(ii)} fast temperature quenches produce ensembles retaining memory of equilibrium ensembles at higher $T$, e.g.\ favoring structures that are more stable against excitation because they lie in deeper energy wells: and \textbf{(iii)} systems exhibit nonexponential relaxation indicative of both glassy dynamics and differing stabilities of degenerate clusters with different structures.
In addition to being directly relevant for experimental studies of small clusters, these results may also improve our understanding of the role such clusters play in controlling kinetic arrest in bulk systems.

\section{Model and Methods}
\label{sec:methods}

The interaction potential for sticky hard spheres with diameter $D$ is \cite{yuste93}:
\begin{equation}
U_{ss}(r) = \Bigg{\{}\begin{array}{ccc}
\infty & , & r < D\\
-\epsilon & , & r = D\\
0 & , & r > D,\\
\end{array}
\label{eq:stickyspherepot}
\end{equation}
where $\epsilon$ is the energy at contact.
A key feature of sticky hard-sphere clusters is that their isoenergetic,
isocontacting states are in general highly degenerate.  
The set of all possible arrangements of $N$ hard spheres with $N_c$ pair contacts consists of $\mathcal{M}(N,N_c)$ nonisomorphic ``macrostates."
$\mathcal{M}$ is an integer for isostatic ($N_c = 3N-6 \equiv N_{ISO}$) and hyperstatic ($N_c > 3N-6$) clusters \cite{hoare76,arkus09} wherein each sphere contacts at least 3
others, and different macrostates have different ``shapes'', i.e.\ distinct sets of interparticle distances $\{r_{ij}^2\}$ ($i,j\in [1,N]$) that correspond to distinguishable inherent structures \cite{stillinger95}.

We determine $\mathcal{M}(N,N_c)$ and find the structure of each macrostate using an updated version of the numerical procedure described at great length in Ref.\ \cite{hoy12}.
The main differences are that here:  \textbf{(a)} we consider adjacency matrices $\{\bar{A}\}$ of arbitrary rather than ``polymeric'' topology;
\textbf{(b)} rather than performing a sequential \cite{hoy12} pass over \textit{all} distinct $\{\bar{A}\}$, we (following Arkus et.\ al.\ \cite{arkus09,arkus11}) use \textit{NAUTY} \cite{mckay13} to generate complete sets of nonisomorphic $\{\bar{A}\}$.
Note that \textbf{(a)} precludes the possibility of failing to detect clusters that do not possess Hamiltonian paths, and that implementing \textbf{(b)} yields an orders-of-magnitude decrease in the computer time (relative to that reported in \cite{hoy12}) required to perform exact enumeration of $\mathcal{M}(N,N_c)$.

\begin{figure}[htbp]
\includegraphics[width=3in]{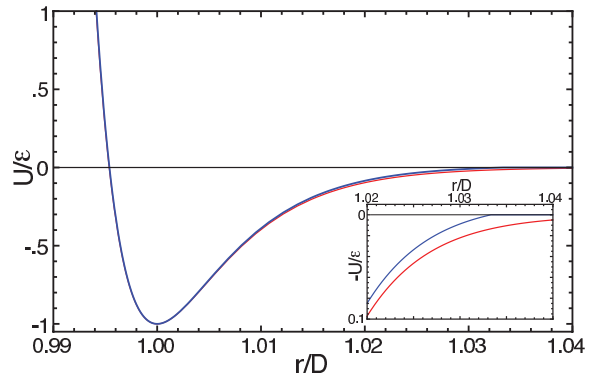}
\caption{Standard (red) and modified (blue) Morse potentials for $\alpha D = 150$.  The inset highlights differences between $U_{\rm Morse}$ and $U_{MM}$ for $r \simeq r_c$.  For $r \lesssim r^{*} \simeq 1.005D$, $U_{MM}$ and $U_{\rm Morse}$ are essentially indistinguishable.}
\label{fig:potential}
\end{figure}

Systems interacting via sticky-hard-sphere potential (Eq.\ \ref{eq:stickyspherepot}) are well known to exhibit anomalous thermodynamics \cite{stell91,foffi00}.
In order to simulate the $T$-dependent structure and dynamics of ``model'' (but realistic) colloidal clusters, a continuous and finite-ranged interaction potential must be introduced.
We perform MD simulations using a modified Morse potential $U_{MM}(r)$ with shape and range (Figure \ref{fig:potential}) similar to the effective interactions between colloids in systems with micellar depletants \cite{meng10,perry12};
\begin{equation}
U_{MM}(\alpha, b; r) = \Bigg{\{}\begin{array}{ccc}\displaystyle\frac{U_{\rm Morse}(\alpha; r) - c(\alpha,b)}{1+c(\alpha,b)} & , & r \leq r_c(\alpha,b)\\
& & \\
0 & , & r > r_c(\alpha,b).
\end{array}
\label{eq:modifiedMorse}
\end{equation}
The structure and dynamics of Morse clusters with large $\alpha D$ are contact-dominated \cite{calvo12}.
In particular, rearrangements can be understood in terms of contact breaking and reformation.
However, defining ``contact'' is ambiguous for potentials that decrease smoothly to zero.
One advantage of using $U_{MM}(r)$ rather than $U_{\rm Morse}(r)$ is that it facilitates contact identification and concomitant analyses of transitions between macrostates; $F_{MM}(r) = -dU_{MM}/dr$ remains finite at $r_c$, allowing us to define contact as finite-force interaction.
The shift/stretch term $c(\alpha, b)$ is defined to make $U_{MM}$ continuous at $r=r_c$, i.e.\ $c(\alpha,b) \equiv U_{\rm Morse}(\alpha; r_c(\alpha, b))$.
We define $b$ to produce a well controlled approximation in which $lim_{b\to\infty} c(\alpha, b) = 0$ and hence 
$lim_{b\to\infty} U_{MM}(\alpha, b; r) = U_{\rm Morse}(\alpha; r)$.
Choosing $r_c(\alpha, b)/D = 1 + b(r^{*}-1)$, where the attractive force $|dU_{\rm Morse}/dr|$ is maximal at 
$r^{*}/D =  (\alpha + \log(2))/\alpha$, gives $c(\alpha,b) = -[4^{-b}(2^{b+1}-1)]\epsilon$.

Here we study systems with $\alpha D=150$ and $b=\alpha/(30\log(2))$, yielding $r_c(\alpha,b)/D = 31/30$. 
We have verified both that this $U_{MM}$ is long-ranged enough to avoid the thermodynamic and dynamic anomalies that are known to arise in the $\alpha \to \infty$ ``Baxter'' limit \cite{stell91}, and that replacement of $U_{\rm Morse}(150;r)$ with this $U_{MM}$ has minimal effects on the structural and dynamic properties of interest here.
Our results should thus be scalable to both larger $\alpha$ and smaller $\alpha$ using (for example) the ``geometrical'' free energy landscape techniques of Holmes-Cerfon et.\ al.\ \cite{holmes13} or the Noro-Frenkel extended law of corresponding states \cite{noro00}.
A preliminary attempt at applying the latter method is reported in the Appendix.

Another advantage of using this $U_{MM}(r)$ is that it allows us to use well-defined ``ideally prepared ensembles'' (IPE) as initial conditions for our MD simulations.
We define IPE as follows:
Suppose a given potential has $\mathcal{M}(N)$ strain-free, energetically degenerate $N$-particle ground state clusters (GSC) with permutational entropies $\omega_k$.
Statistical mechanics predicts that the equilibrium population fraction of each GSC at $T=0$ is $\omega_k/\Omega$, where
\begin{equation}
\Omega = \Omega(N) \equiv \sum_{i=1}^{\mathcal{M}(N)} \omega_k(N).
\label{eq:omegaipe}
\end{equation}
An IPE is an ensemble of molecules containing all of (and only the) $\mathcal{M}(N)$ GSC, such that the population fraction of every GSC is equal to $\omega_k/\Omega$.
Our exact-enumeration studies yield the structures of these GSC \cite{footalphaconv}; values of $\omega_k$ are obtained by evaluating the symmetry of their associated adjacency matrices \cite{arkus11}.
We use IPE of $N_m = f(N) \Omega(N)$ $N$-particle molecules as initial ($T=0$) conditions for MD.  
Here $f(N)$ is chosen to be sufficiently large to give good statistics yet sufficiently small for computational tractability; for the $N=13$ systems studied below we employ $f(13)=1/12972960$, yielding $N_m(13) = \Omega(13)/12972960 = 1290$.
Using $f(N) \neq 1$ simply corresponds to multiplying a system's partition functions by a constant; its value should not (apart from statistical error) alter any results.

MD simulations are performed using an in-house code that employs per-cluster parallelization.
All particles have mass $m$ and diameter $D$.
Each cluster is confined to a cubic cell with hard reflecting walls and side length $L(N)$ chosen to give a particle number density $\rho$ in the dilute limit \cite{malins11}: here $\rho = N/L^{3}(N) = .01D^{-3}$.  
Thus while all particles in a given molecule interact via $U_{MM}(r)$, different clusters do not interact with each other.
This choice of simulation protocol and boundary conditions is motivated by the experiments \cite{meng10,perry12}, which also examined ensembles of isolated $N$-colloid systems in dilute solution.  
MD integration is performed using the velocity-Verlet algorithm with a timestep $\delta t = .03\tau/\alpha$, where the unit of time is $\tau = \sqrt{mD^2/\epsilon}$ \cite{foottimestep}.
Temperature is controlled using a strong Langevin thermostat (with damping time $\tau_{Lang} = \tau$) that mimics the strong damping experienced by colloids in a solvent.  
Comparing to experimental values $\epsilon \simeq 0.1\textrm{eV}$, $D=1\mu m$, and $m \simeq 10^{-15}kg$ \cite{perry12} gives $\tau \simeq 10^{-4}s$.
Our simulations extend as long as $2.5\cdot10^{5}\tau$; this maps to $25s$, which is comparable to the duration of a typical experiment \cite{perry12}.
In Section \ref{sec:results}, all energies, times, and temperatures are respectively expressed in units of $\epsilon$, $\tau$, and $\epsilon/k_B$.

To set up our studies, IPEs are heated from $T = 0$ to $T = 2.5$ (i.e.\ well above the melting point) at a rate $\dot{T}_h^{eq} = 10^{-5}/\tau$.
``Snapshots'' from this heating run are taken at various $T = T_i$ and are further equilibrated at these $T_i$; these equilibrated samples are used as initial conditions for our studies of fixed-$T$ dynamics.
We characterize dynamical relaxation phenomena at these $T$ by examining the traversal of clusters through their various GSCs using the function
\begin{equation}
f_{mad}(t) = \displaystyle\frac{1}{N_m}\sum_{j=1}^{N_m}\left< R(F_j(t'), F_j(t''-t');t',t'') \right>.
\label{eq:fmad}
\end{equation}
Here $f_{mad}(t)$ is the probability that a randomly chosen cluster will \textit{not} execute a transition to a different macrostate within a time interval $t$.  
It is calculated by tracking the structure of each cluster over an ``experimental'' time interval $t = t''-t'$, and then averaging results over all clusters and all ``start times'' $t'$.
In Equation \ref{eq:fmad}, $F_j(t''')$ is the index of the macrostate in which the \textit{j}th  cluster resides at time $t'''$. 
The self-correlation function $R(A,B; t',t'') = 1$ if $A=B$ for \textit{all} internediate times $t'''$ between $t'$ and $t''$, and zero otherwise.  
Thus $f_{mad}(t)$ decays monotonically from one to zero as the ensemble of clusters transition out of their initial states.  

Preparation-protocol-dependence studies are performed by taking the $T=2.5$ end state of the heating run, running at $T=2.5$ for a period of $10^{4}\tau$ in order to obtain a thoroughly equilibrated high-$T$ fluid state, and cooling systems back to $T=0$ at three rates: $|\dot{T}| = 10^{-3}/\tau,\ 10^{-3}/\tau$, and $10^{-5}/\tau$.
During these cooling runs we monitor such quantities as the potential energy of clusters 
\begin{equation}
U = \sum_{i=1}^{N-1} \sum_{j=i+1}^{N} U_{MM}(r_{ij})
\label{eq:UofT}
\end{equation}
and the population fractions of clusters $F_k/N_m$ that correspond to each GSC.  
The latter are identified by comparing their adjacency matrices (assuming particles $i,j$ contact if $r_{ij} < r_c$) to those of the $N_c = N_c^{max}$ packings.
In all cases, finite-$T$ structures correspond either to exactly one zero-temperature GSC, or to an excited state with $N_c < N_c^{max}$.

\section{Results}
\label{sec:results}

The bulk ground states of the sticky-hard-sphere potential (Eq.\  \ref{eq:stickyspherepot}) are the (infinitely degenerate) set $S$  formed by all possible stackings of perfect hexagonal planes into  a close-packed crystal. 
Local ordering within these states may be FCC, HCP, or mixed FCC/HCP.
\textit{Barlow} packings \cite{barlow1883}  are finite-$N$ ``grains'' (subsets) of any member of $S$.
Since they correspond to ``on-pathway'' nuclei that can grow into defect-free members of $S$, they are expected \cite{hoy12} to be be critical to understanding crystal nucleation and growth in systems with hard-core-like repulsions and short-range attractions.
It is important to find all such nuclei that \textit{can} form (as opposed to those that \textit{do} form under specific conditions); this is most conveniently achieved via exact enumeration of sticky-hard-sphere packings.

\begin{figure}[htbp]
\centering
\includegraphics[width=3in]{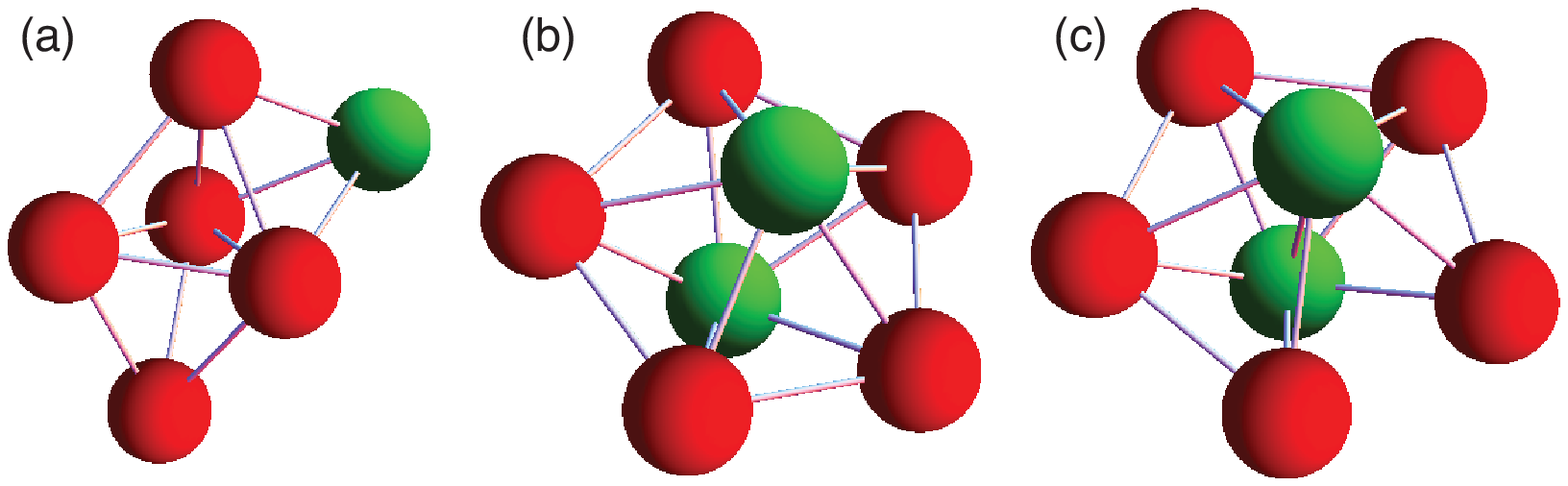}
\centering
\includegraphics[width=2.75in]{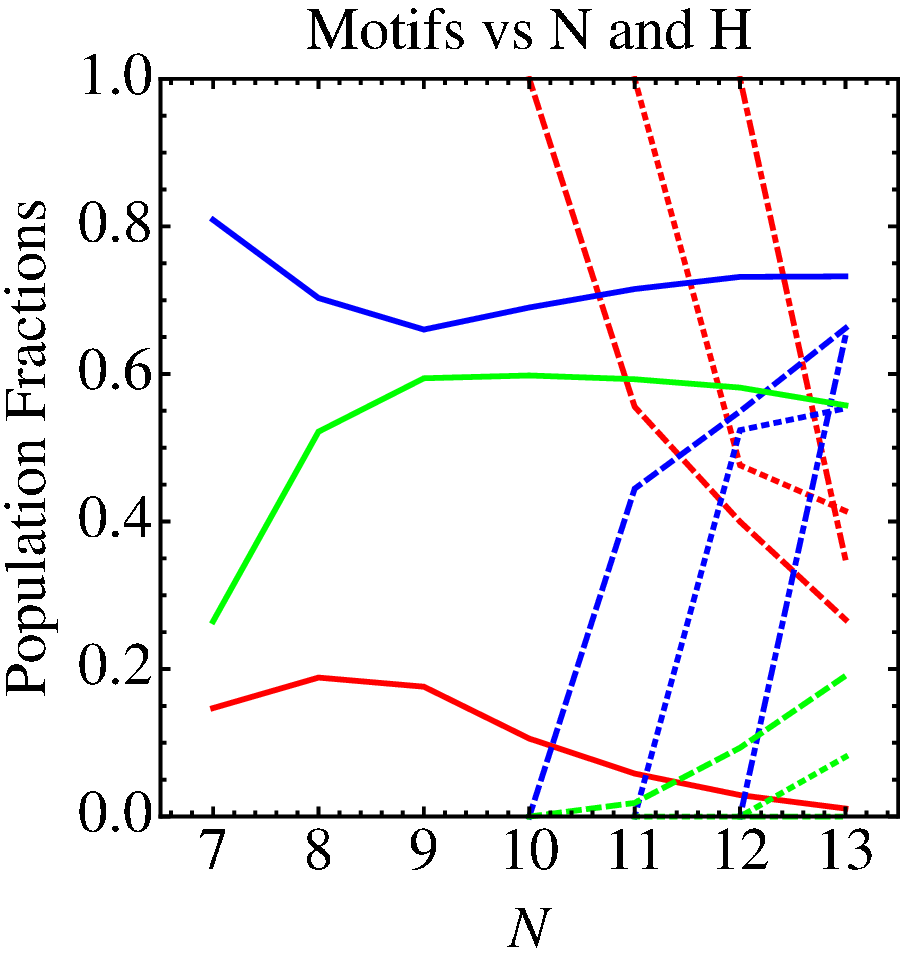}
\caption{Ordered and disordered motifs in monodisperse SHS packings.  (Top) A stack-faulted structure (a), and two fivefold-symmetric structures (b-c).  (Bottom) The population fractions of packings with Barlow order ($f_{Barlow}$; red), fivefold order ($f_{fivefold}$; green), and stacking faults ($f_{stack-fault}$; blue).  Line types are solid for isostatic, dashed for $H=1$, dotted for $H=2$, and dash-dotted for $H=3$.}
\label{fig:motifs}
\end{figure}

In Table \ref{tab:summarytab}, we report the total number of macrostates $\mathcal{M}$, as well as the the numbers of macrostates $\mathcal{M}_X$ possessing structural features $X$ such as Barlow order, stacking faults, and five-fold symmetric defects.
The latter three structural motifs are shown in Figure \ref{fig:motifs}(a-c), preclude Barlow order, and thus correspond to ``off-pathway'' nuclei incompatible with close-packed crystallization.
Here
\begin{equation}
\mathcal{M}_{X}(N,N_c) = \sum_{k=1}^{\mathcal{M}(N,N_c)} G_k(X),
\label{eq:patternM}
\end{equation}
where $G_k(X)$ is $1$ if structure of the $k^{th}$ macrostate matches the pattern $X$ and $0$ otherwise.

\begin{table}[htbp]
\caption{Numbers of macrostates $\mathcal{M}$, macrostates with Barlow order $\mathcal{M}_{Barlow}$, stack faults $\mathcal{M}_{stack-fault}$, and fivefold-symmetric substructures $\mathcal{M}_{fivefold}$. 
Results include both mechanically stable and floppy packings.
Stable packings correspond to zero-dimensional points in configuration space.  Floppy packings occupy finite ``volumes'' in configuration space \cite{holmes13}, but we have verified that all reported here are disconnected from one another, and thus are ``macrostates'' as defined above.  Results for for $N \leq 11$ were reported in Ref.\ \cite{hoy12}, and values of $\mathcal{M}$ agree with those reported in Ref.\ \cite{holmes14}.}
\begin{ruledtabular}
\begin{tabular}{lccccc} 
$N$ & $N_c$ & $\mathcal{M}$ & $\mathcal{M}_{Barlow}$ & $\mathcal{M}_{stack-fault}$ & $\mathcal{M}_{fivefold}$\\
12 & 30 & 11638 & 339 & 8420 & 6657\\
12 & 31 & 174 & 77 & 88 & 16\\
12 & 32 & 8 & 4 & 4 & 0\\
12 & 33 & 1 & 1 & 0 & 0\\
13 & 33 & 95799 & 1070 & 69897 & 53265\\
13 & 34 & 1318 & 363 & 859 & 248\\
13 & 35 & 96 & 42 & 46 & 8\\
13 & 36 & 8 & 5 & 3 & 0
\end{tabular}
\end{ruledtabular}
\label{tab:summarytab}
\end{table}

We find that the fraction of macrostates possessing Barlow order increases rapidly with increasing hyperstaticity $H=N_c-N_{ISO}$, where isostatic packings have $N_{ISO}=3N-6$ contacts.
However, for the range of $N$ considered here, many packings retain non-Barlow order for $H$ as large as three.
Many of these possess stacking faults; $\mathcal{M}_{stack-fault}$ decreases with increasing $H$ but remains nonzero for $H$ up to three.
Fivefold-symmetric motifs are highly prevalent in isostatic packings, and while their prevalence decreases rapidly with increasing $H$, they are still relevant motifs in these more-stable, lower-energy nuclei.

The abovementioned trends are further reinforced by considering the fractions $f_{X}$ of \textit{microstates} with these motifs:
\begin{equation}
f_{X}(N,N_c) = \Omega^{-1} \sum_{k=1}^{\mathcal{M}(N,N_c)} \omega_k G_k(X),
\label{eq:patternf}
\end{equation}
where $\{\omega\}$ and $\Omega$ are given by Equation\ \ref{eq:omegaipe}.
Note that $f_{X}$ is the fraction of clusters in IPEs possessing motif $X$.
Figure \ref{fig:motifs} (bottom panel) shows $f_{Barlow}$, $f_{stack-fault}$ and $f_{fivefold}$ for $7 \leq N \leq 13$ and $0 \leq H \leq 3$.
Notably, $f_{Barlow}$ for isostatic nuclei decreases monotonically with increasing $N$ to only about $1\%$ for $N=13$.
This means that $99\%$ of the highest-energy mechanically stable $N=13$ nuclei are off-pathway, and nucleation of structures with Barlow order is likely to be a rare event.
While $f_{Barlow}$ is far higher for hyperstatic ($H > 0$) nuclei, the same trend of decrease with increasing $N$ persists.

Most non-Barlow nuclei possess stacking faults or fivefold defects; for isostatic nuclei with $8 \leq N \leq 13$, $f_{stack-fault}$ and $f_{fivefold}$ are in the $50-80\%$ range.  
While they decrease sharply with increasing $H$, they still increase in hyperstatic systems to large values with increasing $N$.
Both stack-faulted and fivefold symmetric structures are known to play key roles in inhibiting crystallization in bulk particulate systems by promoting dynamical arrest and glass formation \cite{frank52,royall08}.
Since the energy barriers for transitions between off-pathway and Barlow-ordered nuclei are generally large  \cite{holmes13,morgan14}, the very low values of $f_{Barlow}$ and high values of $f_{stack-fault}$ and $f_{fivefold}$ reported here provide a potential microscopic explanation for the propensity of sticky-hard-sphere-like systems to jam and glass-form in both simulations and experiments (e.g.\ \cite{foffi00,royall08,lu08}).

\begin{figure*}[htbp]
\centering
\includegraphics[width=5.5in]{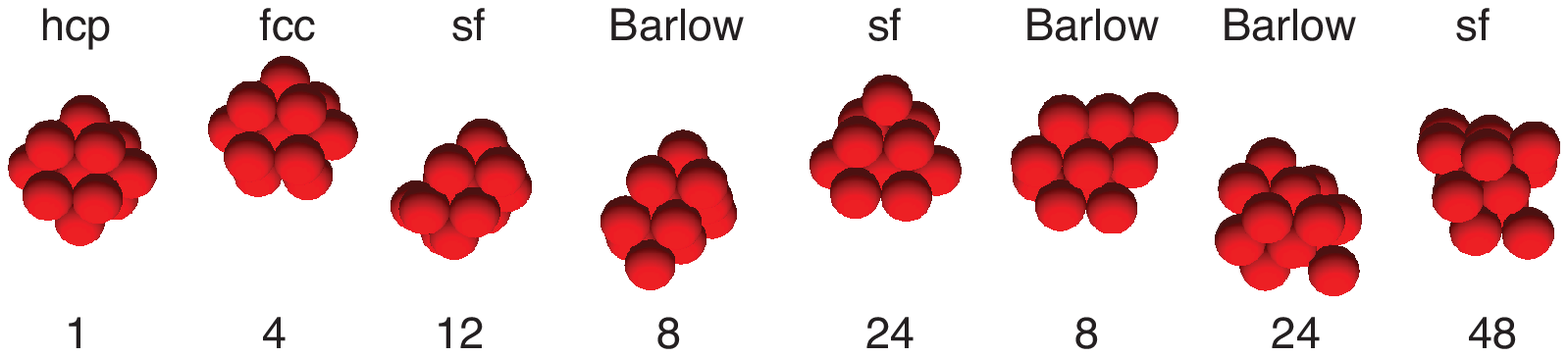}
\centering
\begin{subfigure}
  \centering
\includegraphics[height=1.87in]{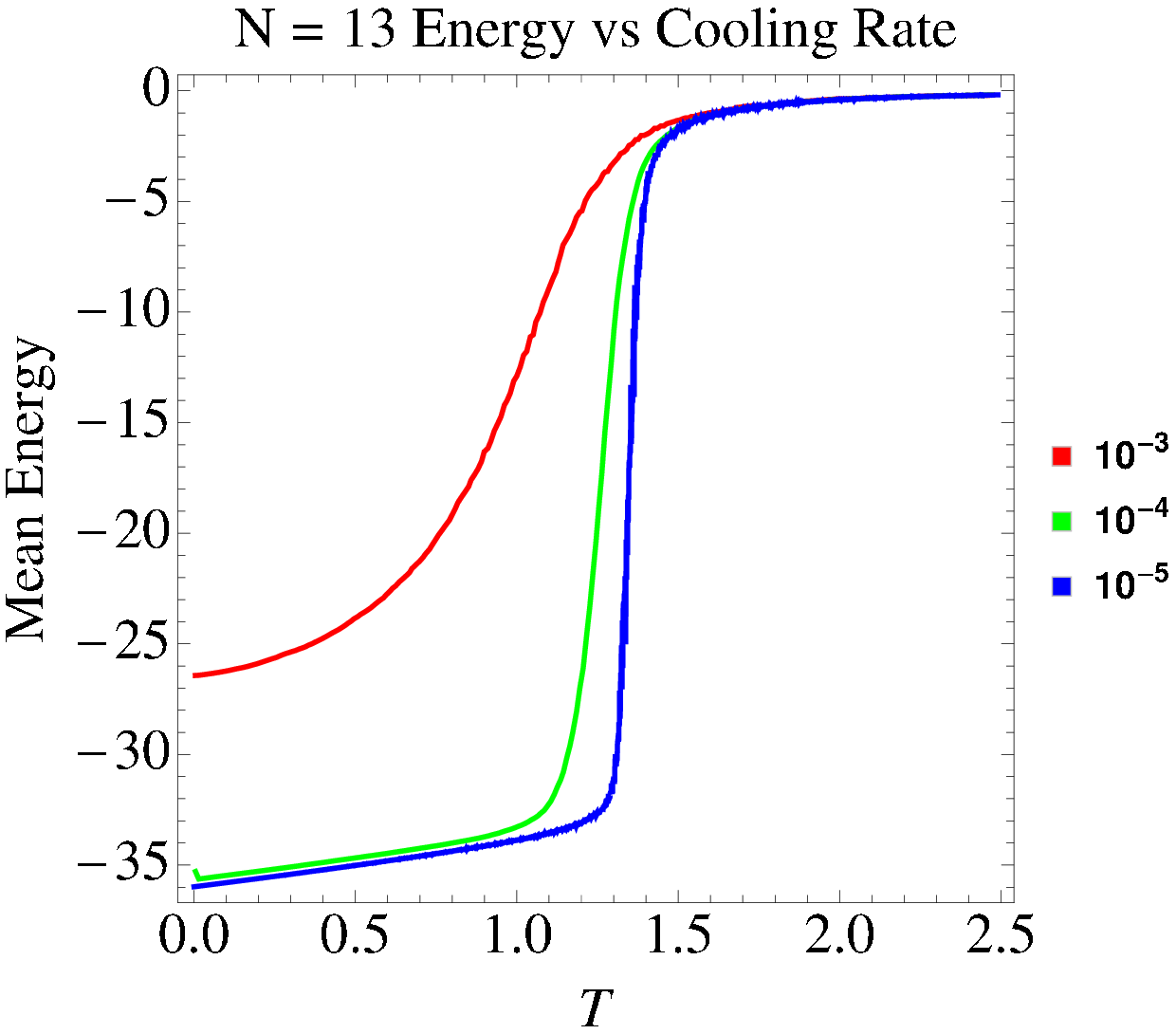}
\end{subfigure}
\begin{subfigure}
  \centering
\includegraphics[height=1.87in]{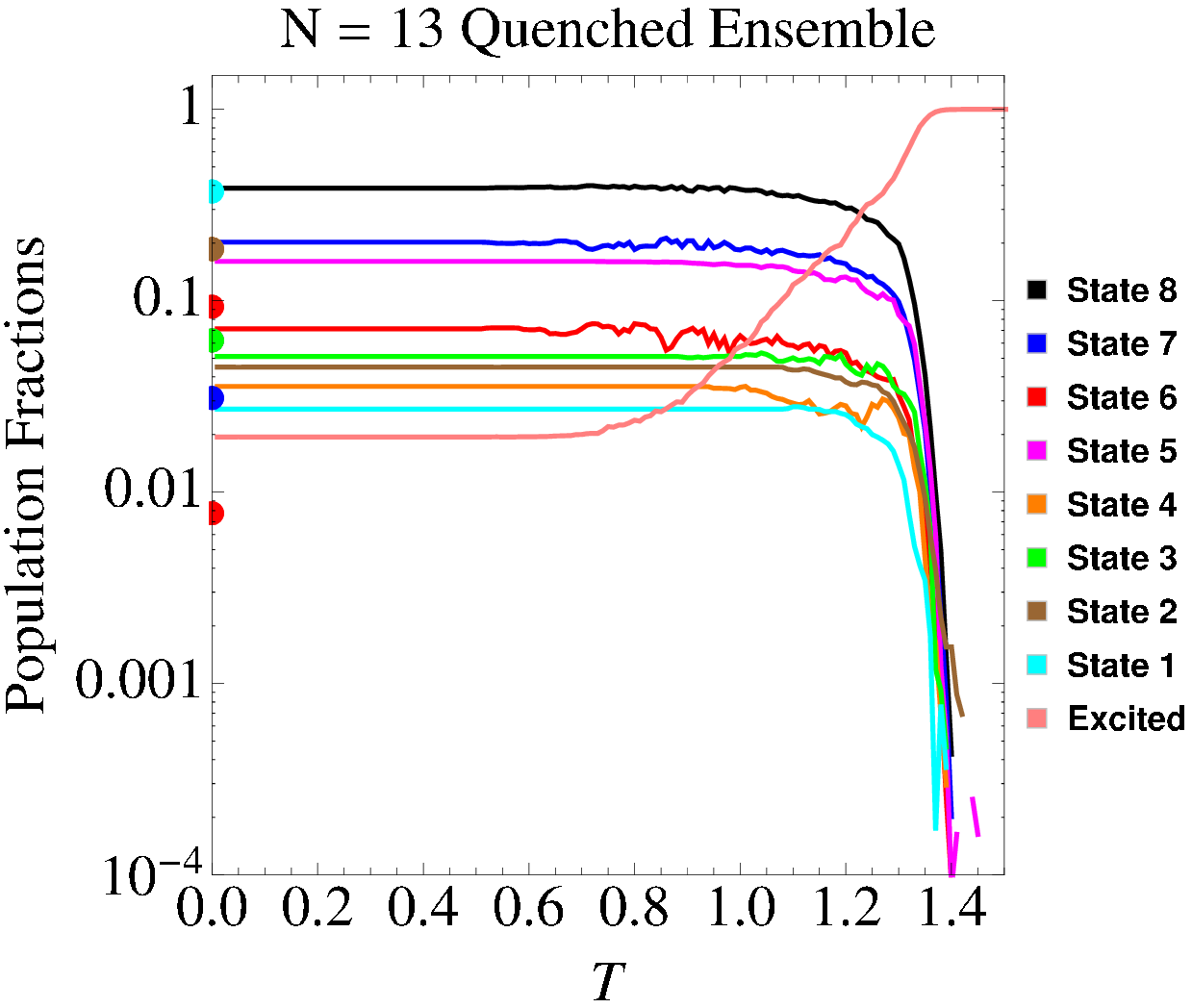}
\end{subfigure}
\begin{subfigure}
  \centering
\includegraphics[height=1.87in]{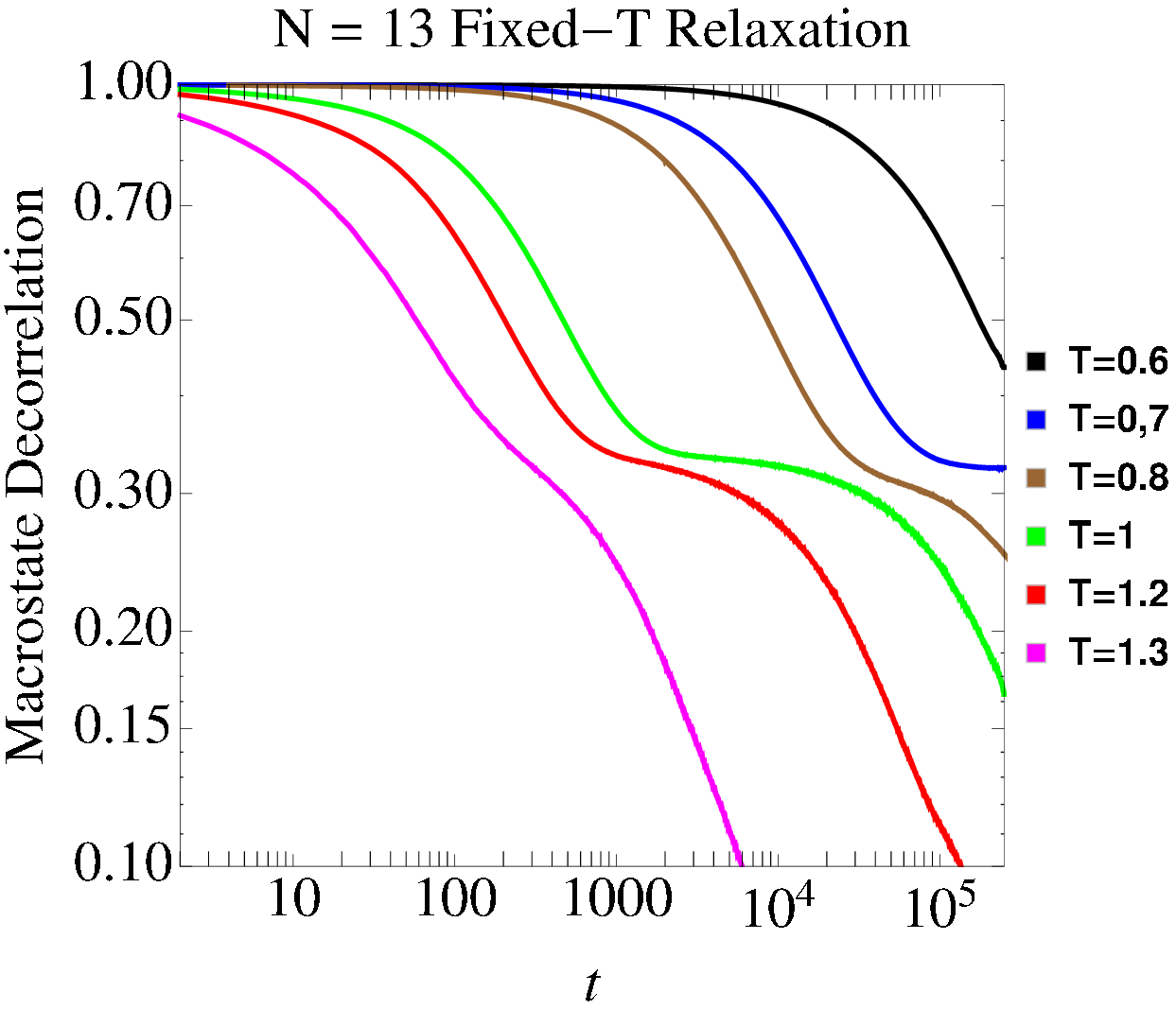}
\end{subfigure}
\caption{Top: The eight $N=13,\ N_c = 36$ ground states.  States 1-8 are depicted from left to right.  Labels above the structures indicate ordering (FCC, HCP, Barlow, or stack-faulted) and numbers below them indicate their relative permutational entropies (ratios of their $\omega_k$).  Bottom: Results from MD simulations for $N=13$ Morse clusters: (Left) Mean cluster energy vs. reduced temperature for fast, medium, and slow quench rates from top to bottom.  (Middle) The population fractions ($F_i/N_m$) of ground states (1-8) and excited states ($(1-\sum_{i=1}^8 F_i)/N_m$) over the course of a slow quench (here $F_i$ is the number of clusters in macrostate $i$.).  Line colors match those of the circles on the left edge, which indicate these states' equilibrium population fractions ($\omega_i/\Omega$) at $T=0$.  (Right) Temperature dependence of cluster relaxation dynamics in thermodynamic equilibrium, as measured by the macrostate decorrelation $f_{mad}(t)$ (Eq.\ref{eq:fmad}).}
\label{fig:1336}
\end{figure*}

In the remainder of this paper, we focus on $N=13$ clusters, and in particular on their nucleation and growth during cooling from high $T$ to $T=0$, as well as on their relaxation dynamics at fixed $T$.
The top panel of Figure \ref{fig:1336} shows the eight degenerate GSCs for $N=13$.
Two are core-shell structures (respectively HCP- and FCC-ordered) wherein a single center sphere contacts twelve neighbors, and the rest are irregularly shaped Barlow and stack-faulted clusters.
Labels above the structures indicate ordering (FCC, HCP, Barlow, or stack-faulted) and numbers below them indicate their relative permutational entropies (ratios of their $\omega_k$) in the IPE.

The left-bottom panel of Figure \ref{fig:1336} shows results for the evolution of the average molecular energy $\left<U(T)\right>$ during cooling from $T=2.5$ to $T=0$ with quench rates $|\dot{T}|$ that vary over a factor of 100.
Results for all $|\dot{T}|$ fall on a common curve above $T_{melt} \simeq 1.5$ since the high-$T$ dynamics are very fast.
Below $T_{melt}$, $\left<U\right>$ begins to drop, indicating the onset of cluster formation.
For the lower two $|\dot{T}|$,  as $T$ continues to decrease, $\left<U\right>$ drops sharply as clusters grow and merge, then flattens out as particles coalesce into single clusters.
A narrow range of small $\left<\partial^2 U/\partial T^2\right>$ indicates a $T$ regime where clusters have coalesced bur continue exploring their energy landscape via inter-macrostate transitions.  
 $\left<\partial U/\partial T\right>$ converges as cluster rearrangement ceases and clusters proceed down the harmonic basins of their energy landscapes.
However, $\left<U\right>/\epsilon$ remains above $-36 = -N_c^{max}$ even at $T=0$, indicating that many clusters freeze into mechanically stable excited states rather than GSCs.  
Results for the fastest quench rate ($10^{-3}$) are markedly different: $\left<U(T)\right>$ decreases much more gradually and remains well above $-(N_{ISO})\epsilon$ even at $T=0$, indicating that systems often freeze into multiple clusters (that do not merge by the end of the cooling runs) rather than single clusters \cite{footmulti}.

The middle-bottom panel shows the population fractions of the GSCs and of excited states as a function of $T$ during the $|\dot{T}| = 10^{-5}/\tau$ quench.
Even at this low cooling rate, about $2\%$ of clusters remain in (mechanically stable) excited states at $T=0$.
The left edge of this panel compares the values of $F_i/N_m$ at the end of the quench to their equilibrium $T=0$ counterparts ($\omega_i/\Omega$ from the IPE).
The FCC and HCP clusters populate the quenched ensemble in excess at low $T$ because they form at slightly higher $T$, and as described below, rearrange more slowly.
Conversely, the other clusters' populations are somewhat lower than equilibrium predictions, showing that for this slow quench rate, clusters inhabiting deep, narrow wells on the potential energy landscape are favored, that is, on-pathway crystal growth is favored.

Higher quench rates (not shown) reverse these trends. 
Clusters are more likely to freeze into less-ordered states that are favored at high $T$ because of their larger vibrational entropy \cite{meng10}, and deviations of the final population fractions from equilibrium $T=0$ values are much larger.

To understand these results, it is useful to recall that the key parameter controlling the growth of ordered crystalline nuclei is the ratio of the particle attachment rate $r_a$ to the cluster reorganization rate $r_r$ \cite{crocker10}.
When $r_a/r_r$ is large, the larger entropy \cite{meng10,hoy12} of disordered (yet mechanically stable) nuclei lacking close-packed order should promote growth of amorphous clusters.  
Conversely, when $r_a/r_r$ is small, enthalpy should rule, and close-packed nuclei should experience stable growth.

Our results are consistent with and reinforce these ideas.
For our fastest quenches, systems often freeze into multiple clusters because $|\dot{T}| > r_a$ even at high $T$.
In contrast, for $|\dot{T}| = 10^{-5}/\tau$, the sharp, first-order-like drop in $\left<U(T)\right>$ is characteristic of the $|\dot{T}| \ll r_a$ regime where single clusters form within a narrow range of $T \simeq T_{melt}$, and the rest of this curve is consistent with $|\dot{T}|$ remaining above $r_r$ down to the $T$ at which $\left<\partial U/\partial T\right>$ converges.
Results in the middle-bottom panel illustrate how $r_r$ grows with decreasing $T$ and increases well beyond $|\dot{T}|$ at $T\simeq 0.6$.

Understanding how $r_r$ varies with $T$ and macrostate is one key to developing principles for controlled \textit{nonequilibrium} self-assembly of these systems.
Towards this end, we now turn to examining their equilibrium relaxation dynamics.
The right-bottom panel of Figure \ref{fig:1336} shows results for the decorrelation $f_{mad}(t)$ of macrostates via state-to-state transitions (Eq.\ \ref{eq:fmad}).
Results are shown for a range of temperatures over which characteristic $r_r$ vary by several orders of magnitude.
At high $T$, excitations from GSCs are very common, energy barriers are easily overcome, and relaxation is nearly exponential.
As $T$ decreases, clear shoulders develop in $f_{mad}(t)$, and relaxation becomes very clearly non-exponential.
One reason for this is that different GSCs possess different stability (i.e.\ lie in potential energy wells of different depths), and so decay at different rates, i.e.\ possess different $r_r$.  
Highly ordered $N=13$ clusters such as HCP and FCC nuclei (states 1-2) are most stable, and have the lowest $r_r$, because every atom in these clusters is bonded to at least five others.  
In contrast, states 7-8 have a ``loose'' atom possessing only three bonds, and rearrange much faster. 
Another potential reason for the complex shapes of $f_{mad}$ observed at lower $T$ is that short-ranged Morse clusters possess glassy dynamics \cite{calvo12}; this will be further examined in forthcoming work.

\section{Discussion and Conclusions}
\label{sec:conclude}

In this paper, we characterized the equilibrium and prepraration-protocol-dependent structure and dynamics of small clusters interacting via hard-core-like repulsions and short-range attractions.
Our results provide a theoretical framework for extending recent experimental studies \cite{malins09,meng10,perry12} of small colloidal clusters to examine both equilibrium relaxation dynamics at fixed $T$ and a variety of nonequilibrium phenomena.
In particular, they should be relevant to understanding the factors controlling \textit{nonequilibrium} self-assembly of such clusters, and should be testable using plausible extensions of currently available experimental techniques \cite{perry12,fung12,fung13}.

We extended recent exact enumeration studies of sticky hard sphere packings \cite{arkus09,hoy10,arkus11,hoy12} to $N = 13$. 
This is an important advance because $N=13$ clusters can form complete core-shell structures (i.e.\ HCP and FCC crystallites); our work will aid experimental studies of core-shell structures where observation of the inner-core particles is difficult.
We then employed these complete sets of packings as ``ideally-prepared-ensemble'' (IPE) initial conditions for MD simulations of 
colloids interacting via a short-ranged modified Morse potential, focusing on $N=13$ clusters.

It is important to note that the results presented here are strictly valid only for systems interacting via ``steep'' (short-ranged) pair potentials.  
Softer, longer-ranged interactions dramatically alter the lower regions of small clusters' energy landscapes \cite{doye95,morgan14}.  
However, the short-ranged limit considered here is experimentally accessible, e.g.\ in systems of micron-sized colloids and micellar depeletants \cite{perry12}. 
To aid experimental tests of our results, we include an Appendix containing a Noro-Frenkel analysis \cite{noro00} that can be used for mapping them to systems interacting via other pair potentials.

We gratefully acknowledge Miranda Holmes-Cerfon for sharing preliminary results for $N \geq 12$ packings \cite{miranda,holmes14}, and Miranda Holmes-Cerfon, David Wales, and Paddy Royall for helpful discussions.

\begin{appendix}

\section{Noro-Frenkel Analysis}

Our results can be used to make predictions for systems interacting via other short-ranged pair potentials - including experimental systems (see e.g.\ Ref.\ \cite{lu08}) - using Noro and Frenkel's extension \cite{noro00} of the law of corresponding states.  
Both thermodynamical and dynamical results can be effectively compared by ``temperature-matching'' different systems at the same value of ``free volume concentration'' $c_p  = \pi\rho\sigma_{eff}^{3}/6$ and the reduced second virial coefficient
\begin{equation}
B_2^*(T) = \displaystyle\frac{3}{2\sigma_{eff}^3(T)} \int_{0}^{r_c(a,b)} \left[1-\exp{(-U_{MM}(r)/k_BT)}\right] r^2 dr.
\label{eq:B2}
\end{equation}
Here the temperature-dependent effective hard-sphere diameter \cite{andersen71} is
\begin{equation}
\sigma_{eff}(T) = \int_{0}^{1} \left[1-\exp{(-U_{MM}(r)}\right] dr.
\label{eq:seff}
\end{equation}

Values of $\sigma_{eff}(T)$ and $B_2^*(T)$ for the temperatures examined in the lower-right panel of Fig.\ \ref{fig:1336} are given in Table \ref{tab:B2tab}. 
The variation of $B_2^*$ with $T$ is small because for the steep, short-ranged interaction potential $U_{MM}$ used in this study, the integrand in Eq.\ \ref{eq:B2} is close to unity except in a very narrow range $\delta r \sim (r_c - 1)$ about $r=1$.
However, our study of dynamical relaxation in equilibrium systems suggests that the timescales as well as the character of relaxation in real systems with similarly short-ranged interactions  can vary very sharply over a narrow range of $B_2^*$.  
Future work will consider wider ranges of $N$, $c_p$ and $B_2^*$ in order to allow comparison to published results for phenomena such as dynamical arrest in individual clusters \cite{kroy04,malins09} and bulk systems \cite{lu08}, as well as guiding future experiments.

\begin{table}[htbp]
\caption{Values of $\sigma_{eff}(T)$ and $B_2^*(T)$ (Eqs.\ \ref{eq:B2}-\ref{eq:seff}) for the temperatures examined in the lower-right panel of Fig.\ \ref{fig:1336}.}
\begin{ruledtabular}
\begin{tabular}{lcc} 
$T$ & $\sigma_{eff}$ & $B_2^*$\\
0.6 & 0.9824 & 0.9046\\ 
0.7 & 0.9853 & 0.9267\\ 
0.8 & 0.9871 & 0.9407\\
1.0 & 0.9891 & 0.9572\\
1.2 & 0.9901 & 0.9666\\
1.3 & 0.9905 & 0.9699\\
\end{tabular}
\end{ruledtabular}
\label{tab:B2tab}
\end{table}

\end{appendix}


\begin{thebibliography}{38}
\bibitem[{foo({\natexlab{a}})}]{footalphaconv}
\bibinfo{note}{This procedure is simple only for potentials with short-ranged
  attractive tails. For example, $U_{MM}$ is sufficiently short-ranged for
  $\alpha > \alpha_{conv}$, where $\alpha_{conv}$ as the interaction range such
  that $r_c(\alpha, b)$ is greater than all GSC's minimum
  second-nearest-neighbor distance $r_2$ for $\alpha < \alpha_{conv}$ and less
  than $r_2$ for $\alpha < \alpha_{conv}$. For $N \leq 13$, $\alpha_{conv}(N) <
  30$.}
\bibitem[{foo({\natexlab{b}})}]{foottimestep}
\bibinfo{note}{We have verified that this value of $\delta t$ is sufficiently
  small by checking that the velocity autocorrelation function is insensitive
  to varying $\delta t$ over the range $[.002\tau/\alpha,.007\tau/\alpha]$ in
  systems at $k_B T/\epsilon = 1$.}
\bibitem[{foo({\natexlab{c}})}]{footmulti}
\bibinfo{note}{The $\left<U(T)\right> > -N_{ISO}\epsilon$ condition may be
  understood by examining the maximum number of pair contacts possessed by
  multiple clusters. For example, two clusters of $N_1$ and $N_2$ atoms with
  $N_1 + N_2 \leq 24$ can form no more than $N_c^{max}(N_1) + N_c^{max}(N_2) =
  3N-6 = N_{ISO}$ contacts \cite{arkus11}.}
\end{thebibliography}

\end{document}